\documentclass[journal]{IEEEtran}

\usepackage[utf8]{inputenc}
\usepackage[T1]{fontenc}
\usepackage{cite}
\usepackage{amsmath,amssymb,amsfonts}
\usepackage{algorithm}
\usepackage{algpseudocode}
\usepackage{graphicx}
\usepackage{textcomp}
\usepackage{xcolor}
\usepackage{booktabs}
\usepackage{url}
\usepackage{multirow}
\usepackage{array}
\usepackage{siunitx}
\usepackage{hyperref}
\usepackage{stfloats}
\usepackage{placeins}
\usepackage{float}
\usepackage{enumitem}

\setlist{noitemsep, topsep=2pt, partopsep=0pt, parsep=0pt, leftmargin=*}

\begin{document}

\title{A Multi-Agent System for 5G Throughput Prediction in Multi-Operator Urban Environments}

\author{Muhammad~Kabeer\textsuperscript{\,1,2},~\IEEEmembership{Student Member,~IEEE,}
        Rosdiadee~Nordin\textsuperscript{\,3,4},~\IEEEmembership{Senior Member,~IEEE,}
        Nadiva~Nuriftitah\textsuperscript{\,1},
        and~Sian~Lun~Lau\textsuperscript{\,1},~\IEEEmembership{Senior Member,~IEEE}
\thanks{\textsuperscript{1}\,Faculty of Engineering and Technology, Sunway University, Selangor, Malaysia.}
\thanks{\textsuperscript{2}\,Department of Computer Science, Federal University Dutsinma, Katsina, Nigeria.}
\thanks{\textsuperscript{3}\,Future Cities Research Institute (FCRI), Faculty of Engineering and Technology, Sunway University, No. 5, Jalan Universiti, Bandar Sunway, 47500 Selangor Darul Ehsan, Malaysia.}
\thanks{\textsuperscript{4}\,Future Cities Research Institute (FCRI), Lancaster University, Lancaster, LA1 4YW, United Kingdom.}
\thanks{Corresponding author: R. Nordin (email: rosdiadeen@sunway.edu.my).}}

\maketitle

\begin{abstract}
Throughput prediction is foundational for artificial intelligence-driven 6G resource orchestration. Conventional monolithic machine learning models struggle to generalize across diverse operators, mobility modes, and traffic types, leaving a critical stochasticity gap between signal conditions and achievable throughput. To overcome these constraints in heterogeneous urban environments, we propose a Tiered Multi-Agent System (TMAS) that dynamically routes edge telemetry to context-aware Domain Micro-Agents, validated on a dataset of 48,618 samples collected in Sunway City, Malaysia, with Nemo Handy drive test software, spanning three Tier-1 mobile network operators, three mobility modes, namely (i) elevated pedestrian walkway, (ii) ground-level shuttle bus, and (iii) elevated bus rapid transit; and three traffic profiles, namely (i) persistent download, (ii) persistent upload, and (iii) adaptive video streaming. Our evaluations reveal that TMAS overcomes predictability bottlenecks, achieving a coefficient of determination ($R^{2}$) of up to 0.931 and a Mean Absolute Error (MAE) as low as 0.53 Mbps. The system demonstrates high operational efficiency, with rapid micro-agent training times, low inference latencies, and agentic routing overhead of 0.004 to 0.126 ms. These latency characteristics indicate the architecture is a promising candidate for the response times required by next-generation wireless networks.
\end{abstract}

\begin{IEEEkeywords}
Throughput Prediction, Ensemble Learning, Mobility Awareness, AI-Native Networks, Multi-Agent System.
\end{IEEEkeywords}

\section{Introduction}
\IEEEPARstart{T}{he} global rollout of 5G New Radio (NR) has initiated a paradigm shift in wireless communication, moving beyond simple mobile broadband toward enhanced mobile broadband, ultra-reliable low-latency communication, and massive machine-type communications \cite{s19173651}, critical for data-intensive services such as autonomous vehicular networks, remote surgery, and immersive 8K video streaming. However, the efficacy of these services is intrinsically tied to the network's ability to maintain high quality of service in dynamic environments, motivating accurate, real-time throughput prediction for proactive resource allocation, seamless mobility management, and intelligent edge orchestration.

Traditional throughput estimation has relied on statistical averaging or reactive measurements that fail to capture rapid cellular fluctuations. Early work achieved high predictability under stable, line-of-sight conditions \cite{11278580, 9495144}, but as 5G transitions into modern cities, predictability is compromised by severe multipath fading, waveguide effects in complex structures, human body blockage at millimeter-wave frequencies, and dual-connectivity fluctuations \cite{11170415, 10.1145/3636534.3694725, 10147378}. Beyond the physical layer, adaptive bitrate (ABR) logic used by streaming services such as YouTube introduces buffer-driven traffic patterns decoupled from instantaneous signal strength \cite{HELMY2025104324, 10770241}, necessitating a transition from universal modeling to context-aware frameworks that account for specific mobility modes and application types.

Despite the surge in deep-learning-based solutions, a trade-off persists between precision, computational overhead, and cross-domain generalization. Monolithic models and standard federated learning architectures assume a degree of data homogeneity that does not exist across competing Tier-1 operators, whose non-independent and identically distributed (non-i.i.d.) data silos cause centralized models to suffer weight divergence, while complex deep architectures often exceed the 10 ms decision window required for real-time edge deployment \cite{HASAN2026100721}. Ensemble methods such as XGBoost and Random Forest have shown promise in capturing nonlinear throughput fluctuations and coverage patterns \cite{10007850, 10418223}, yet none of these studies quantify the predictability gaps introduced by distinct urban mobility modes or application layer ABR logic, nor propose a decentralized routing architecture for heterogeneous operator data, the central contribution of this work.

This work addresses three interlocking gaps: environmental fading within the modern city of Sunway City, application layer buffer decoupling, and operator heterogeneity. The primary contributions of this research are:

\begin{enumerate}
\item A high-fidelity \href{https://data.mendeley.com/datasets/8vvn3k3hmz/1}{dataset} \cite{kabeer2026multioperator} of 48,618 samples captured with \href{https://www.keysight.com}{Keysight's} Nemo Handy drive test software across three Malaysian operators, three mobility modes, and three traffic profiles in Sunway City, Malaysia, enabling systematic evaluation of operator, mobility, and application effects on throughput predictability.
\item TMAS, a three-tier Mixture-of-Experts architecture comprising a Master Agent for context orchestration, Operator Agents for silo isolation, and Domain Micro-Agents for specialized prediction, dynamically routing inference requests to circumvent the weight divergence that plagues monolithic models on non-i.i.d.\ data.
\item A systematic evaluation of TMAS across application, mobility, and operator dimensions, including inference latency and training time benchmarks confirming that TMAS micro-agents meet the fast response times required by software-defined networking and next-generation intelligent controllers.
\end{enumerate}

The remainder of this paper is organized as follows. Section II reviews related work. Section III details the measurement campaign and the TMAS architecture. Section IV presents empirical results. Section V discusses implications for 6G and core orchestration. Section VI concludes.

\section{Related Work}
The research trajectory in 5G NR throughput prediction has shifted from static physical layer models toward multi-dimensional, context-aware approaches that integrate mobility and service dynamics, reviewed critically below.

\subsection{Empirical Characterization of Environmental Dynamics}
A fundamental challenge in throughput prediction is the non-stationary nature of the wireless channel. Foundational work, such as the Salzburg Research campaign by Dabarera et al.\ \cite{11278580}, established predictive baselines using multi-operator data along stable highway segments, achieving high deterministic accuracy in open topographies but often failing to account for modern city dynamics. In contrast, several studies have mapped the severe multipath fading and waveguide effects that dominate dense transit structures: Zhao et al.\ \cite{11170415} and Feng et al.\ \cite{10.1145/3636534.3694725} characterize tunnel and mmWave environments, while Geo2ComMap \cite{10950386} instead leverages geographic data to predict MIMO throughput. These studies establish that environmental context, particularly mobility mode and topography, shapes the predictability ceiling, yet none systematically quantify how different mobility regimes alter model performance, a gap this work fills with per-mobility performance metrics across three distinct urban modes.

\subsection{Methodological Evolution}
The methodology for network key performance indicator prediction has undergone a significant transition. Early frameworks relied on low-complexity linear approaches, including multiple linear regression with Kalman filtering \cite{10342799}, Bayesian and kernel regression \cite{11059833}, and legacy or load-balancing models \cite{Al-Thaedan2023, 10914824}, all of which fail to capture the rapid fading of 5G. The field subsequently moved toward nonlinear temporal architectures, including LSTM and transformer-based models for flow, traffic-heatmap, and bandwidth-delay forecasting \cite{9878077, LI2025111669, 10.1145/3703629}, multimodal and CNN-memory fusion networks \cite{11167476, 11274662}, WLAN-focused transformers \cite{11317760, 10992709}, and end-to-end deep learning pipelines \cite{10971899}. Despite their precision, these deep models often exceed the sub-10 ms decision windows required for real-time edge processing, motivating a shift toward ensemble and gradient-boosting methods \cite{electronics11081227} that offer a superior latency-accuracy Pareto front; Yuliana et al.\ \cite{10418223} and Fazwan and Nordin \cite{10007850} both identify RF and XGBoost as top performers for coverage and throughput shifts, respectively. Critically, however, these ensemble-based studies have not evaluated whether their reported accuracy holds across distinct mobility regimes or application types.

\subsection{Application Layer Stochasticity and ABR Dynamics}
A critical stochasticity gap exists between physical signal parameters and actual application-layer throughput. Studies on ABR streaming \cite{HELMY2025104324, 10770241} demonstrate that video buffer logic introduces autonomous traffic patterns decoupled from instantaneous RSRP or signal-to-interference-plus-noise ratio; researchers have addressed this disconnect through edge-assisted and energy-efficient streaming robustness \cite{11141395, 10056411}, human-centric quality-of-experience mapping \cite{10192432}, closed-loop bandwidth clustering \cite{10929655}, and multistage prediction of application-layer capping \cite{10.1145/3651863.3651878, 10.1145/3724400}. High-level quality of experience cannot be managed by universal signal models alone. Prior work, however, has not quantified how much predictability degrades moving from controlled download traffic to adaptive streaming under identical radio conditions; this study provides that quantification, revealing the severe predictive degradation introduced by YouTube's ABR logic and establishing a concrete stochasticity gap that must be bridged by application-aware modeling.

\subsection{Heterogeneity and Cross-Domain Generalization}
The modern 5G landscape is characterized by carrier aggregation and dual connectivity, where independent carrier prediction \cite{10.1145/3651890.3672250}, self-attention weighting of disparate signals \cite{10147378}, online model adaptation \cite{10.1145/3724400}, and federated learning \cite{10838521, 10.1145/3709141} have each been proposed to manage model aging and privacy during telemetry collection. However, these works share a common limitation: generalization failure across operational domains, as models optimized for one mobility mode or operator infrastructure degrade substantially when transferred to another due to differences in tower density, scheduling algorithms, and frequency allocations. Threshold selection techniques in cognitive radio networks \cite{RATHORE2026111824} can optimize throughput in high-traffic ecosystems, but require additional signaling that may not be available across operators. This study systematically quantifies these cross-domain generalization gaps, providing empirical evidence that context-aware, granular modeling is not merely beneficial but necessary for urban 5G deployment.

Collectively, prior work evaluates ensemble and deep learning models in controlled settings but rarely spans multiple mobility modes, operators, and application layers simultaneously. This study fills that gap by benchmarking Universal Fallback architectures against Domain Micro-Agents across three mobility modes, three operators, and three traffic profiles within a single dense urban environment.

\section{Methodology}
\begin{figure*}[t]
\centering
\includegraphics[width=0.78\textwidth]{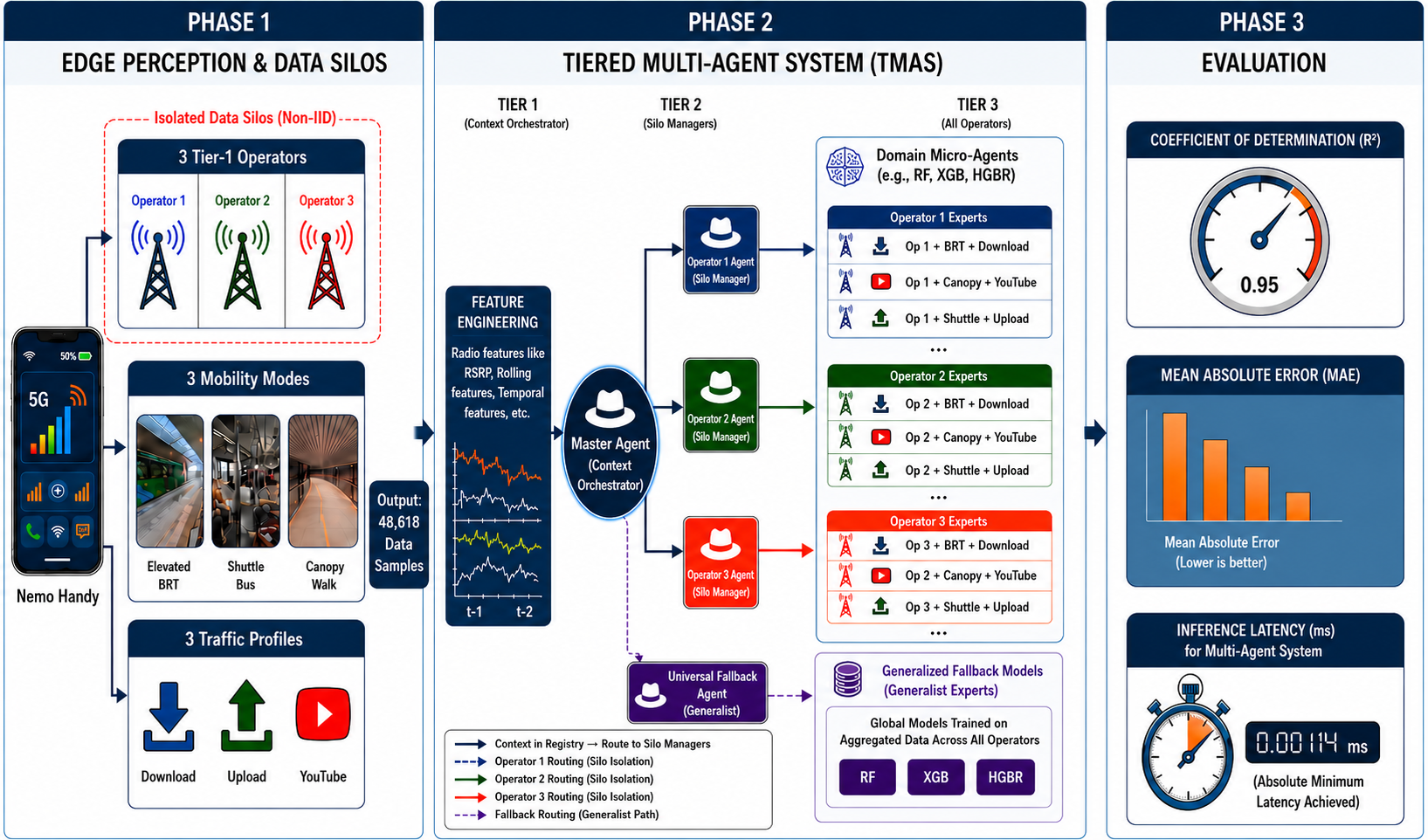}
\caption{The proposed TMAS architecture: Phase 1 performs edge perception across independent operator silos, Phase 2 shows dynamic routing from the Master Agent through Operator Agents to Domain Micro-Agents, and Phase 3 evaluates execution latency.}
\label{fig:method}
\end{figure*}

To resolve the latency and data divergence limitations of monolithic modeling, this research proposes the TMAS architecture \cite{HASAN2026100721}, a tiered, data-driven framework for characterizing and predicting 5G NR throughput within a non-stationary urban environment. As illustrated in Fig.~\ref{fig:method}, the architecture is structured into three primary phases: edge perception of isolated data silos, hierarchical agentic routing, and closed-loop evaluation.

\subsection{Phase 1: Edge Perception and Data Silos}

The measurement campaign was localized in Sunway City, Malaysia, a modern city characterized by high-rise structures, elevated walkways, and mixed line-of-sight and non-line-of-sight propagation conditions. To capture the diverse signal propagation signatures of the city's topography, three distinct mobility modes were orchestrated.

The mobility profiles include:
\begin{itemize}
    \item \textbf{Canopy Walk:} Elevated pedestrian walkways involving low-velocity mobility, dense urban foliage, and complex structural reflections.
    \item \textbf{Shuttle Bus:} Ground-level feeder routes characterized by severe metal body shadowing and multipath interference from high-rise structures.
    \item \textbf{Elevated Bus Rapid Transit (BRT):} A high-velocity environment on an elevated guideway, providing a mix of line-of-sight and rapid fading zones.
\end{itemize}

Data was acquired across three Tier-1 Malaysian mobile network operators, anonymized as Operator 1, Operator 2, and Operator 3 to protect proprietary network configurations. Each combination of operator, mobility mode, and traffic profile was collected over multiple drive test passes. No samples were excluded based on throughput extremes, and GPS coordinates were retained without outlier removal since they are integral to spatial context. Because disparate operator infrastructures possess distinct scheduling algorithms and tower densities, they are treated as isolated data silos. Telemetry was captured using a professional drive test tool to gather high-fidelity physical layer data without requiring inter-operator data aggregation.

The acquisition targeted three traffic profiles: (i) persistent download, (ii) persistent upload, and (iii) adaptive video (YouTube 4K), yielding a consolidated dataset of 48,618 samples at a 1-second temporal resolution; the resulting per-profile sample composition is summarized in Table~\ref{tab:data_stats}. The upload profile is comparatively under-represented, reflecting a subset of operator-mobility combinations that experienced repeated session dropouts due to application-layer socket interruptions and uplink scheduler starvation, which prevented sufficient sample accumulation for those specific contexts. The remaining operator-mobility-traffic combinations completed logging successfully across all three passes. Key extracted radio features include RSRP, temporal lags ($\text{RSRP}_{\text{Rolling5}}$, $\text{RSRP}_{\text{lag}}$), and spatial coordinates (latitude, longitude). Additional features include the physical cell identifier and the absolute radio frequency channel number.

\begin{table}[ht]
\caption{Summary of Dataset Throughput Statistics (Mbps)}
\label{tab:data_stats}
\centering
\begin{tabular}{@{}lrrrrr@{}}
\toprule
\textbf{Category} & \textbf{Samples} & \textbf{Mean} & \textbf{Std. Dev.} & \textbf{Median} & \textbf{Max} \\ \midrule
\textbf{Traffic Profile} \\
Download & 24,020 & 4.19 & 4.85 & 2.46 & 30.50 \\
Upload & 7,421 & 18.05 & 11.80 & 21.04 & 45.44 \\
YouTube & 17,177 & 14.83 & 24.71 & 3.32 & 364.17 \\ \midrule
\textbf{Mobility Mode} \\
BRT & 11,941 & 10.91 & 16.93 & 2.57 & 201.97 \\
Shuttle & 14,324 & 9.01 & 11.21 & 3.17 & 79.20 \\
Canopy & 22,353 & 10.28 & 19.53 & 2.48 & 364.17 \\ \bottomrule
\end{tabular}
\end{table}

\subsection{Phase 2: Tiered Multi-Agent System}
To circumvent the computational bloat and domain generalization failures of universal modeling, the predictive architecture is decomposed into a three-tier routing system (Fig.~\ref{fig:method}, Phase 2), formalized in Algorithm~\ref{alg:tmas}. Each sample is assigned a discrete context key $\kappa \triangleq (o, m, u) \in \mathcal{O} \times \mathcal{M} \times \mathcal{U}$, identifying its operator, mobility mode, and traffic profile, with $D_{\kappa} \subseteq D$ denoting the corresponding contextual subset of the dataset $D$. Offline, the pipeline trains a universal baseline $\mathcal{F}^{*}$ from candidate models $f \in \{\text{RF, XGB, HGBR}\}$ on the full dataset, and, for every context whose subset satisfies $|D_{\kappa}| \geq \tau_{\min}$, trains a specialized Domain Micro-Agent on the entire contextual subset $D_{\kappa}$, storing the best-performing model in the registry $\mathcal{A}[\kappa]$. The threshold $\tau_{\min}$ serves only as a minimum-viability floor below which a context is deemed too data-sparse to support reliable model fitting; it is not the training sample size itself. Once a context clears this floor, the corresponding micro-agent is trained on all available samples for that operator-mobility-traffic combination, which for most contexts numbers in the hundreds to thousands. At inference, the Master Agent routes an incoming feature vector $\mathbf{x}$ with context $\kappa$: it dispatches to the matching micro-agent $f \leftarrow \mathcal{A}[\kappa]$ if $\kappa \in \mathcal{A}$, or defaults to the universal fallback $f \leftarrow \mathcal{F}^{*}$ otherwise, to compute the prediction $\hat{y} = f(\mathbf{x})$.

The routing hierarchy proceeds as follows:

\begin{itemize}
    \item \textbf{Tier 1 (Master Agent):} Performs rule-based context extraction from the operator, mobility, and traffic profile flags, routing to the matching Silo Manager if the context is registered, or to the Universal Fallback Agent otherwise.
    \item \textbf{Tier 2 (Silo Managers):} Routes registered contexts to the corresponding Operator Agent, keeping predictions within the correct operator's domain and avoiding weight divergence from non-i.i.d.\ operator data.
    \item \textbf{Tier 3 (Domain Micro-Agents):} Delegates the final prediction to a lightweight gradient boosting ensemble mapped to the exact operator-mobility-traffic combination (e.g., Op 1 + BRT + Download), preserving minimal inference latency by activating only the required micro-model.
\end{itemize}

\begin{algorithm}[t]
\caption{TMAS: Training and Agentic Routing}
\label{alg:tmas}
\begin{algorithmic}[1]

\Require Dataset $D$ with context key $\kappa = (o,m,u) \in \mathcal{O} \times \mathcal{M} \times \mathcal{U}$;
         minimum-viability threshold $\tau_{\min}$;
         query vector $\mathbf{x}$
\Ensure  Universal model $\mathcal{F}^{*}$; micro-agent registry $\mathcal{A}$;
         prediction $\hat{y}$ with provenance label $\ell$

\Statex \textbf{Stage 1: Training (offline)}
\State Train $f \in \{\text{RF, XGB, HGBR}\}$ on $D$; set $\mathcal{F}^{*} \leftarrow$ best model
\For{each context $\kappa = (o, m, u) \in \mathcal{O} \times \mathcal{M} \times \mathcal{U}$}
    \State $D_{\kappa} \leftarrow$ \textsc{Filter}$(D, \kappa)$
    \If{$|D_{\kappa}| \geq \tau_{\min}$}
        \State Train RF, XGB, HGBR on $D_{\kappa}$; store best model as $\mathcal{A}[\kappa]$
    \EndIf
\EndFor

\Statex \textbf{Stage 2: Agentic Routing (online)}
\Function{RouteAndPredict}{$\mathbf{x}$}
    \State $\kappa \leftarrow$ \Call{MasterAgent.ExtractContext}{$\mathbf{x}$} \Comment{Tier 1: context orchestration}
    \If{$\kappa \in \mathcal{A}$}
        \State $\text{OpAgent} \leftarrow$ \Call{SiloManager.Resolve}{$\kappa.o$} \Comment{Tier 2: operator silo isolation}
        \State $f \leftarrow$ \Call{OpAgent.Dispatch}{$\mathcal{A}[\kappa]$} \Comment{Tier 3: domain micro-agent}
        \State $\ell \leftarrow$ Domain Micro-Agent
    \Else
        \State $f \leftarrow \mathcal{F}^{*}$;\ \ $\ell \leftarrow$ Universal Fallback
    \EndIf
    \State \Return $f(\mathbf{x}),\, \ell$
\EndFunction

\end{algorithmic}
\end{algorithm}

\subsection{Learning Architectures: RF, HGBR, and XGBoost}
For the Domain Micro-Agents, we evaluate three standard ensemble learners, namely Random Forest (RF), Histogram-based Gradient Boosting Regression (HGBR), and XGBoost (XGB), selected for their established latency-accuracy Pareto front on tabular, non-i.i.d.\ data. Each candidate is trained per context with shallow tree depths and modest ensemble sizes to keep inference within the sub-10 ms edge budget, and the best-performing architecture per context is retained in the registry $\mathcal{A}[\kappa]$.

\subsection{Phase 3: Execution Latency Benchmarking}
To validate the real-time feasibility of the TMAS architecture, Phase 3 quantifies the end-to-end inference latency and training overhead of each candidate micro-agent. Inference latency ($T_{\text{inf}}$) is measured per sample on an edge-class CPU without GPU acceleration to reflect realistic deployment constraints. Training time is recorded for each domain-specific model. These measurements directly inform the trade-off between accuracy and computational cost, providing a practical guide for selecting the optimal micro-agent for each context. The results of this benchmarking are presented in the Complexity Analysis subsection of Section~\ref{sec:results} below.

\subsection{Evaluation Metrics}
Model performance is quantified using the coefficient of determination ($R^2$) and mean absolute error (MAE):
\begin{equation}
\text{MAE} = \frac{1}{n} \sum_{i=1}^{n} |y_i - \hat{y}_i|
\end{equation}
alongside per-sample inference latency ($T_{\text{inf}}$, in ms), the primary benchmark for deployment feasibility in real-time edge systems.

\section{Results and Empirical Analysis}
\label{sec:results}

\subsection{Dataset Characterization}
We first examine the raw throughput distributions to understand the inherent variability across different traffic types. Table~\ref{tab:data_stats} summarizes the throughput statistics across these dimensions.

\begin{figure}[t]
\centering
\includegraphics[width=\columnwidth]{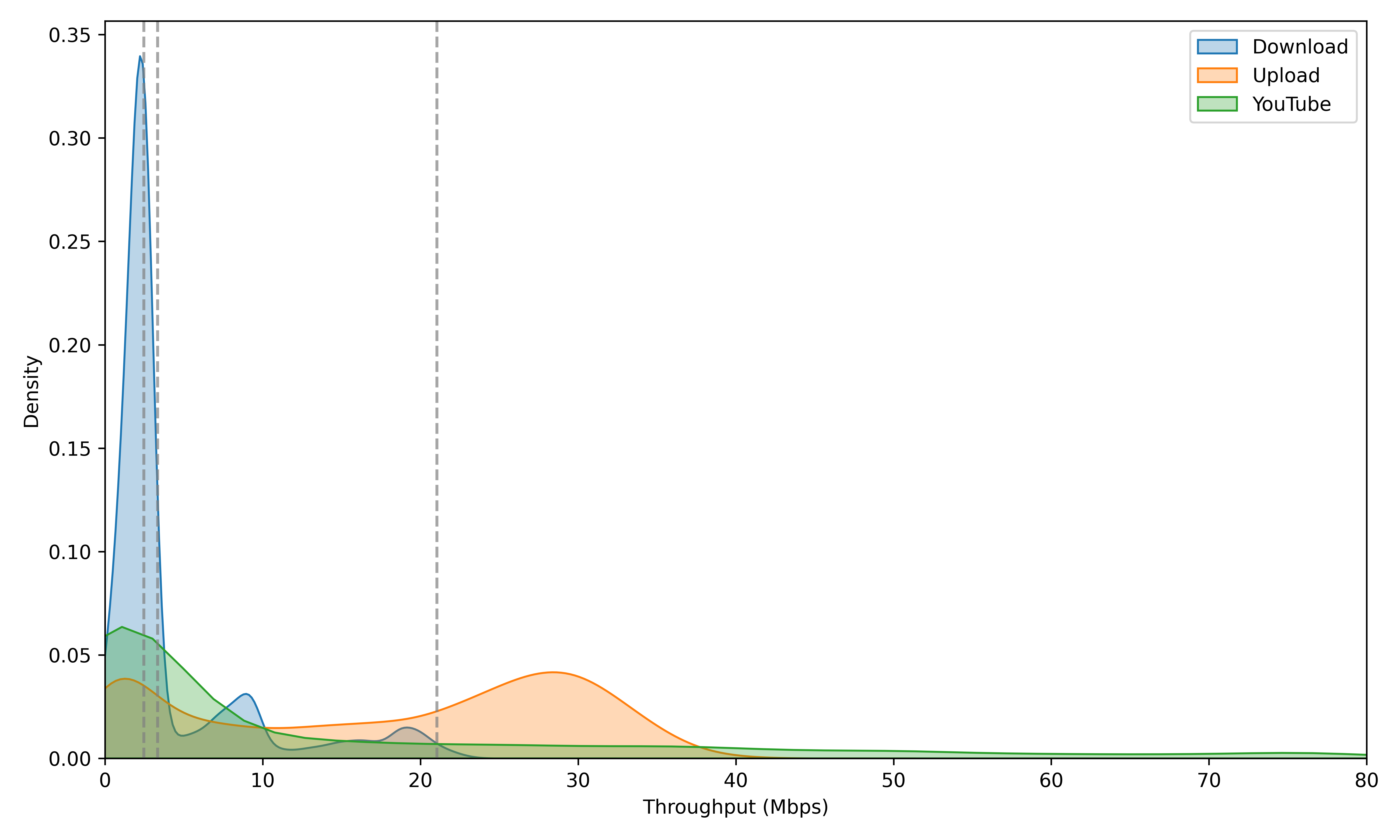}
\caption{Probability density function of throughput at the application layer across all three traffic profiles.}
\label{fig:pdf}
\end{figure}

\begin{figure*}[t]
\centering
\includegraphics[width=0.8\textwidth]{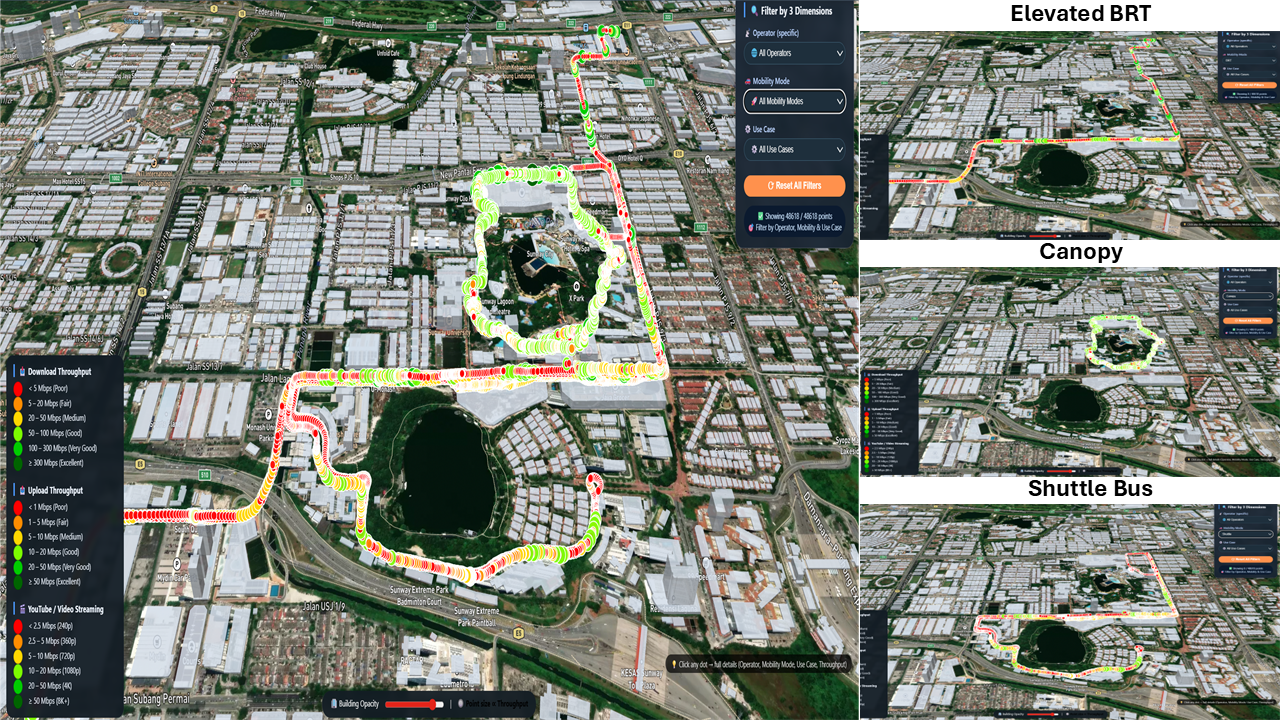}
\caption{Overlay of throughput measurements across the three mobility modes, namely (i) Canopy Walk, (ii) Shuttle Bus, and (iii) Elevated BRT, and the combined dataset.}
\label{fig:measurements}
\end{figure*}

As visualized in Figure~\ref{fig:pdf}, the probability density functions confirm distinct behavioral signatures for download, upload, and YouTube traffic. The download profile is tightly constrained, forming a sharp, prominent spike at the lower end of the spectrum with a standard deviation of 4.85 Mbps. The upload distribution forms a secondary, wider hump near 30 Mbps, reflecting its higher mean throughput of 18.05 Mbps. The presence of long tails is markedly evident in the YouTube category; while it clusters densely near zero, it stretches far outward, illustrating the long-tail bursts and high standard deviation of 24.71 Mbps. Figure~\ref{fig:measurements} provides a spatial overlay of the throughput data collected for each mobility mode.

\subsection{Application-Level Performance}
Table~\ref{tab:generalized_models} presents the comprehensive performance of generalized agents, including universal, traffic-specific, and mobility-specific models across all three architectures. Table~\ref{tab:domain_models} details the performance of the specific domain micro-agents utilizing a grouped layout mapping predictive metrics to operational subsets. The specialized micro-agents substantially outperform the universal baseline. For Download, the best Domain Micro-Agent (Operator 1, RF) attains $R^{2}=0.931$, whereas the monolithic Universal RF reaches only $0.326$. For YouTube, the optimal micro-agent (Operator 1, RF) achieves $R^{2}=0.339$, while the universal XGB gives $0.291$. 

Download dominates the $R^2$ metric across all models, consistently near 0.9, while YouTube results are visibly depressed, yielding the lowest $R^2$ and highest MAE, reflecting operator-specific ABR buffering behavior. This delta validates the TMAS routing logic: forcing a single model to generalize across heterogeneous operator data degrades accuracy, whereas dynamic routing to specialized experts preserves precision.

\subsection{Mobility Mode Performance}
Table~\ref{tab:domain_models} presents the performance of operator-specific domain micro-agents across the three mobility modes. In the $R^2$ results, BRT and Shuttle display notably higher stability across most operators, yielding up to $R^2 = 0.751$ (Operator 1, RF). A notable exception is observed for Operator 1 on the Shuttle Bus route, where the RF model spikes substantially to $R^2 = 0.924$, exceeding BRT performance for the same operator, attributable to the dominance of download traffic samples within the Operator 1 Shuttle dataset, where persistent download conditions produce a more predictable throughput profile that overwhelms the multipath complexity of the ground-level route. Conversely, the Canopy $R^2$ values remain below 0.35 across the board, establishing it as the most difficult environment; the accompanying MAE results invert this pattern, with Canopy presenting the highest error values. This hierarchy reflects the progressive increase in radio environment complexity: BRT benefits from near-line-of-sight conditions, while Shuttle Bus and Canopy suffer from severe multipath, metal-body shadowing, and foliage attenuation. Operator 1 outperforms the other two operators in most mobility mode combinations, whereas Operator 3 generally exhibits the lowest predictability.

\begin{table}[t]
\caption{Performance of Generalized Agents (Mean $\pm$ Std over 3 Runs)}
\label{tab:generalized_models}
\centering
\resizebox{\linewidth}{!}{%
\begin{tabular}{@{}llcccc@{}}
\toprule
\textbf{Category} & \textbf{Model} & \textbf{$R^2$} & \textbf{MAE (Mbps)} & \textbf{Inf. Lat. (ms)} & \textbf{Train Time (s)} \\ \midrule
\multirow{3}{*}{Universal}
& HGBR & $0.278 \pm 0.009$ & $7.91 \pm 0.11$ & $0.00246 \pm 0.00007$ & $1.32 \pm 1.42$ \\
& RF   & $0.326 \pm 0.014$ & $6.64 \pm 0.09$ & $0.00971 \pm 0.00054$ & $3.58 \pm 0.05$ \\
& XGB  & $0.291 \pm 0.006$ & $7.57 \pm 0.07$ & $0.00298 \pm 0.00127$ & $0.61 \pm 0.18$ \\ \midrule
\multirow{3}{*}{Download}
& HGBR & $0.923 \pm 0.003$ & $0.79 \pm 0.01$ & $0.00290 \pm 0.00006$ & $0.42 \pm 0.04$ \\
& RF   & $0.931 \pm 0.004$ & $0.73 \pm 0.01$ & $0.01129 \pm 0.00127$ & $1.67 \pm 0.03$ \\
& XGB  & $0.925 \pm 0.003$ & $0.78 \pm 0.01$ & $0.00351 \pm 0.00032$ & $0.44 \pm 0.07$ \\ \midrule
\multirow{3}{*}{Upload}
& HGBR & $0.595 \pm 0.011$ & $5.24 \pm 0.13$ & $0.00565 \pm 0.00027$ & $0.37 \pm 0.03$ \\
& RF   & $0.646 \pm 0.019$ & $4.70 \pm 0.17$ & $0.03160 \pm 0.00088$ & $0.53 \pm 0.00$ \\
& XGB  & $0.609 \pm 0.017$ & $5.10 \pm 0.21$ & $0.01092 \pm 0.00135$ & $0.42 \pm 0.03$ \\ \midrule
\multirow{3}{*}{YouTube}
& HGBR & $0.160 \pm 0.011$ & $14.94 \pm 0.25$ & $0.00251 \pm 0.00081$ & $0.27 \pm 0.08$ \\
& RF   & $0.234 \pm 0.037$ & $14.16 \pm 0.33$ & $0.01353 \pm 0.00037$ & $1.19 \pm 0.02$ \\
& XGB  & $0.186 \pm 0.039$ & $15.06 \pm 0.28$ & $0.00508 \pm 0.00044$ & $0.45 \pm 0.05$ \\ \midrule
\multirow{3}{*}{BRT}
& HGBR & $0.419 \pm 0.006$ & $6.94 \pm 0.17$ & $0.00452 \pm 0.00028$ & $0.36 \pm 0.01$ \\
& RF   & $0.484 \pm 0.011$ & $5.64 \pm 0.17$ & $0.01974 \pm 0.00023$ & $0.79 \pm 0.00$ \\
& XGB  & $0.457 \pm 0.040$ & $6.61 \pm 0.10$ & $0.00736 \pm 0.00138$ & $0.43 \pm 0.05$ \\ \midrule
\multirow{3}{*}{Shuttle}
& HGBR & $0.453 \pm 0.016$ & $4.52 \pm 0.08$ & $0.00311 \pm 0.00058$ & $0.34 \pm 0.06$ \\
& RF   & $0.447 \pm 0.013$ & $4.08 \pm 0.06$ & $0.01686 \pm 0.00240$ & $0.91 \pm 0.02$ \\
& XGB  & $0.414 \pm 0.005$ & $4.59 \pm 0.10$ & $0.00722 \pm 0.00194$ & $0.47 \pm 0.03$ \\ \midrule
\multirow{3}{*}{Canopy}
& HGBR & $0.208 \pm 0.016$ & $9.60 \pm 0.05$ & $0.00122 \pm 0.00007$ & $0.20 \pm 0.02$ \\
& RF   & $0.225 \pm 0.030$ & $8.56 \pm 0.07$ & $0.01438 \pm 0.00193$ & $1.62 \pm 0.01$ \\
& XGB  & $0.157 \pm 0.022$ & $9.42 \pm 0.09$ & $0.00434 \pm 0.00067$ & $0.46 \pm 0.03$ \\ \bottomrule
\end{tabular}
}
\end{table}

\begin{table*}[t]
\caption{Performance and Efficiency of Operator-Specific Domain Micro-Agents (Mean $\pm$ Std over 3 Runs)}
\label{tab:domain_models}
\centering
\resizebox{\textwidth}{!}{%
\begin{tabular}{@{}llcccccccccccc@{}}
\toprule
& & \multicolumn{3}{c}{\textbf{$R^2$}} & \multicolumn{3}{c}{\textbf{MAE (Mbps)}} & \multicolumn{3}{c}{\textbf{Inference Latency (ms)}} & \multicolumn{3}{c}{\textbf{Train Time (s)}} \\
\cmidrule(lr){3-5} \cmidrule(lr){6-8} \cmidrule(lr){9-11} \cmidrule(lr){12-14}
\textbf{Domain} & \textbf{Model} & \textbf{Operator 1} & \textbf{Operator 2} & \textbf{Operator 3} & \textbf{Operator 1} & \textbf{Operator 2} & \textbf{Operator 3} & \textbf{Operator 1} & \textbf{Operator 2} & \textbf{Operator 3} & \textbf{Operator 1} & \textbf{Operator 2} & \textbf{Operator 3} \\ \midrule
\multirow{3}{*}{Download}
& HGBR & $0.926 \pm 0.012$ & $0.878 \pm 0.010$ & $0.792 \pm 0.007$ & $1.00 \pm 0.05$ & $0.56 \pm 0.02$ & $0.71 \pm 0.01$ & $0.00434 \pm 0.00015$ & $0.00539 \pm 0.00068$ & $0.00487 \pm 0.00024$ & $0.37 \pm 0.05$ & $0.34 \pm 0.01$ & $0.34 \pm 0.01$ \\
& RF   & $0.931 \pm 0.011$ & $0.891 \pm 0.003$ & $0.805 \pm 0.004$ & $0.95 \pm 0.05$ & $0.53 \pm 0.02$ & $0.69 \pm 0.01$ & $0.02198 \pm 0.00048$ & $0.03073 \pm 0.00043$ & $0.03269 \pm 0.00138$ & $0.62 \pm 0.01$ & $0.49 \pm 0.00$ & $0.46 \pm 0.00$ \\
& XGB  & $0.923 \pm 0.013$ & $0.877 \pm 0.009$ & $0.775 \pm 0.014$ & $1.02 \pm 0.04$ & $0.57 \pm 0.02$ & $0.74 \pm 0.01$ & $0.00874 \pm 0.00063$ & $0.01332 \pm 0.00293$ & $0.01400 \pm 0.00567$ & $0.44 \pm 0.02$ & $0.43 \pm 0.05$ & $0.42 \pm 0.03$ \\ \midrule
\multirow{3}{*}{Upload}
& HGBR & $0.374 \pm 0.055$ & $0.710 \pm 0.009$ & $0.498 \pm 0.025$ & $7.08 \pm 0.47$ & $4.32 \pm 0.09$ & $4.26 \pm 0.20$ & $0.01220 \pm 0.00102$ & $0.00658 \pm 0.00052$ & $0.02010 \pm 0.00002$ & $0.31 \pm 0.02$ & $0.33 \pm 0.02$ & $0.25 \pm 0.02$ \\
& RF   & $0.457 \pm 0.048$ & $0.752 \pm 0.019$ & $0.526 \pm 0.031$ & $6.46 \pm 0.56$ & $3.82 \pm 0.15$ & $4.04 \pm 0.22$ & $0.10342 \pm 0.00483$ & $0.04125 \pm 0.00655$ & $0.18904 \pm 0.00601$ & $0.19 \pm 0.01$ & $0.34 \pm 0.00$ & $0.16 \pm 0.00$ \\
& XGB  & $0.410 \pm 0.013$ & $0.711 \pm 0.015$ & $0.481 \pm 0.009$ & $6.75 \pm 0.36$ & $4.31 \pm 0.14$ & $4.25 \pm 0.07$ & $0.04141 \pm 0.00275$ & $0.01874 \pm 0.00684$ & $0.10877 \pm 0.05215$ & $0.36 \pm 0.01$ & $0.43 \pm 0.02$ & $0.38 \pm 0.01$ \\ \midrule
\multirow{3}{*}{YouTube}
& HGBR & $0.199 \pm 0.053$ & $0.180 \pm 0.019$ & $0.115 \pm 0.011$ & $13.38 \pm 0.26$ & $14.08 \pm 0.15$ & $15.70 \pm 0.07$ & $0.00760 \pm 0.00091$ & $0.00594 \pm 0.00039$ & $0.00601 \pm 0.00037$ & $0.36 \pm 0.04$ & $0.33 \pm 0.01$ & $0.33 \pm 0.01$ \\
& RF   & $0.339 \pm 0.058$ & $0.187 \pm 0.019$ & $0.125 \pm 0.021$ & $11.98 \pm 0.18$ & $13.94 \pm 0.23$ & $15.33 \pm 0.07$ & $0.04238 \pm 0.00604$ & $0.03565 \pm 0.00099$ & $0.03486 \pm 0.00026$ & $0.38 \pm 0.01$ & $0.43 \pm 0.01$ & $0.50 \pm 0.02$ \\
& XGB  & $0.142 \pm 0.097$ & $0.105 \pm 0.040$ & $0.023 \pm 0.020$ & $13.91 \pm 0.11$ & $14.53 \pm 0.08$ & $16.42 \pm 0.05$ & $0.01779 \pm 0.00110$ & $0.01784 \pm 0.00503$ & $0.01199 \pm 0.00139$ & $0.46 \pm 0.05$ & $0.40 \pm 0.03$ & $0.45 \pm 0.01$ \\ \midrule
\multirow{3}{*}{BRT}
& HGBR & $0.693 \pm 0.009$ & $0.332 \pm 0.011$ & $0.430 \pm 0.068$ & $3.44 \pm 0.18$ & $9.76 \pm 0.25$ & $7.55 \pm 0.14$ & $0.00684 \pm 0.00086$ & $0.00733 \pm 0.00060$ & $0.00904 \pm 0.00022$ & $0.32 \pm 0.02$ & $0.33 \pm 0.01$ & $0.32 \pm 0.01$ \\
& RF   & $0.751 \pm 0.019$ & $0.390 \pm 0.069$ & $0.538 \pm 0.065$ & $2.78 \pm 0.21$ & $8.42 \pm 0.46$ & $6.43 \pm 0.24$ & $0.03886 \pm 0.00531$ & $0.05706 \pm 0.00019$ & $0.05739 \pm 0.00274$ & $0.39 \pm 0.01$ & $0.30 \pm 0.01$ & $0.21 \pm 0.01$ \\
& XGB  & $0.696 \pm 0.021$ & $0.372 \pm 0.035$ & $0.424 \pm 0.087$ & $3.42 \pm 0.17$ & $9.31 \pm 0.54$ & $7.29 \pm 0.26$ & $0.01572 \pm 0.00228$ & $0.02645 \pm 0.00775$ & $0.02346 \pm 0.00030$ & $0.42 \pm 0.02$ & $0.43 \pm 0.05$ & $0.39 \pm 0.00$ \\ \midrule
\multirow{3}{*}{Shuttle}
& HGBR & $0.920 \pm 0.011$ & $0.244 \pm 0.013$ & $0.720 \pm 0.018$ & $1.32 \pm 0.05$ & $7.24 \pm 0.11$ & $2.65 \pm 0.04$ & $0.00637 \pm 0.00033$ & $0.00587 \pm 0.00032$ & $0.00698 \pm 0.00109$ & $0.33 \pm 0.02$ & $0.33 \pm 0.02$ & $0.32 \pm 0.03$ \\
& RF   & $0.924 \pm 0.005$ & $0.211 \pm 0.009$ & $0.728 \pm 0.030$ & $1.27 \pm 0.05$ & $7.11 \pm 0.07$ & $2.47 \pm 0.06$ & $0.04613 \pm 0.00871$ & $0.03782 \pm 0.00076$ & $0.04226 \pm 0.00318$ & $0.30 \pm 0.02$ & $0.37 \pm 0.01$ & $0.31 \pm 0.01$ \\
& XGB  & $0.917 \pm 0.001$ & $0.151 \pm 0.026$ & $0.745 \pm 0.034$ & $1.34 \pm 0.03$ & $7.59 \pm 0.14$ & $2.53 \pm 0.06$ & $0.01689 \pm 0.00175$ & $0.01483 \pm 0.00275$ & $0.01961 \pm 0.00433$ & $0.41 \pm 0.03$ & $0.43 \pm 0.02$ & $0.44 \pm 0.05$ \\ \midrule
\multirow{3}{*}{Canopy}
& HGBR & $0.267 \pm 0.027$ & $0.311 \pm 0.007$ & $0.121 \pm 0.004$ & $9.74 \pm 0.53$ & $6.19 \pm 0.33$ & $11.87 \pm 0.43$ & $0.00595 \pm 0.00045$ & $0.00509 \pm 0.00051$ & $0.00527 \pm 0.00071$ & $0.31 \pm 0.01$ & $0.33 \pm 0.00$ & $0.34 \pm 0.04$ \\
& RF   & $0.335 \pm 0.076$ & $0.314 \pm 0.011$ & $0.120 \pm 0.022$ & $8.71 \pm 0.77$ & $5.67 \pm 0.34$ & $11.78 \pm 0.41$ & $0.03213 \pm 0.00069$ & $0.02442 \pm 0.00222$ & $0.03251 \pm 0.00080$ & $0.46 \pm 0.01$ & $0.56 \pm 0.02$ & $0.48 \pm 0.02$ \\
& XGB  & $0.241 \pm 0.074$ & $0.238 \pm 0.035$ & $-0.003 \pm 0.013$ & $9.81 \pm 0.73$ & $6.28 \pm 0.41$ & $12.53 \pm 0.35$ & $0.01310 \pm 0.00402$ & $0.00863 \pm 0.00099$ & $0.01161 \pm 0.00240$ & $0.38 \pm 0.02$ & $0.43 \pm 0.02$ & $0.41 \pm 0.05$ \\ \bottomrule
\end{tabular}%
}
\end{table*}

\subsection{Universal Model Scatter Analysis}
\begin{figure*}[t]
\centering
\includegraphics[width=\textwidth]{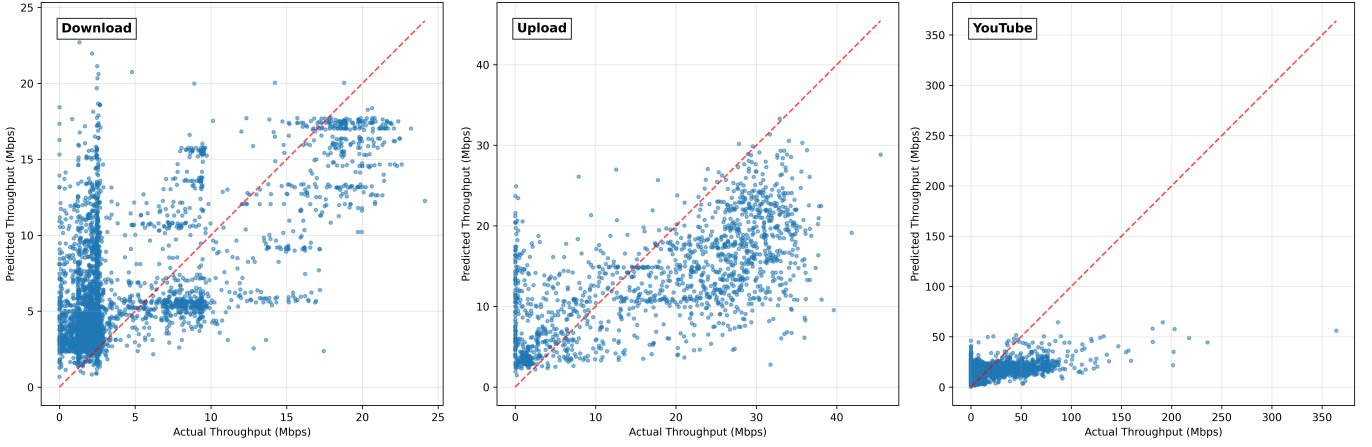}
\caption{HGBR model predictions versus actual throughput for Download, Upload, and YouTube traffic profiles evaluated on the universal model.}
\label{fig:scatter}
\end{figure*}

To illustrate the behavior of a single universal model across different traffic types, we trained a HistGradientBoostingRegressor (HGBR) model on the entire dataset and evaluated it separately on each traffic profile. As shown in Figure~\ref{fig:scatter}, the predicted versus actual throughput reveals distinct behavioral constraints. The Download profile forms a highly dense cluster cleanly tracking along the ideal diagonal red line, indicating robust accuracy. The Upload scatter is more dispersed, forming an under-prediction cloud beneath the diagonal. Critically, the YouTube scatter plot reveals a dense concentration near zero, coupled with a severe horizontal dispersion; actual throughput sporadically spikes up to 364.17 Mbps, but the model predictions rigidly plateau below 60 Mbps, visually demonstrating the model's failure to capture ABR-driven bursts.

\subsection{Feature Importance}
To ensure consistent structural importance metrics across all algorithms, specifically because HGBR does not natively expose Gini impurity arrays, we apply a parallel Random Forest proxy trained on the same data partition to derive feature rankings for each traffic type. This analysis reveals a pronounced uniformity across all three profiles: latitude and longitude are the dominant features, substantially exceeding all other parameters. This indicates that precise geographic positioning, which implicitly encodes static environmental blockages such as high-rise structures and urban foliage, is the primary driver of achievable throughput. Following spatial coordinates, temporal signal metrics, specifically the 5-sample rolling RSRP average ($\text{RSRP}_{\text{Rolling5}}$) and historical lag features, dominate the decision trees, while categorical network features (such as PCI, ARFCN, and Band) rank practically negligible. This demonstrates that the physical channel state and geographic location govern throughput predictability far more than macroscopic network labels.

\subsection{Complexity Analysis}
\begin{figure}[t]
\centering
\includegraphics[width=\columnwidth]{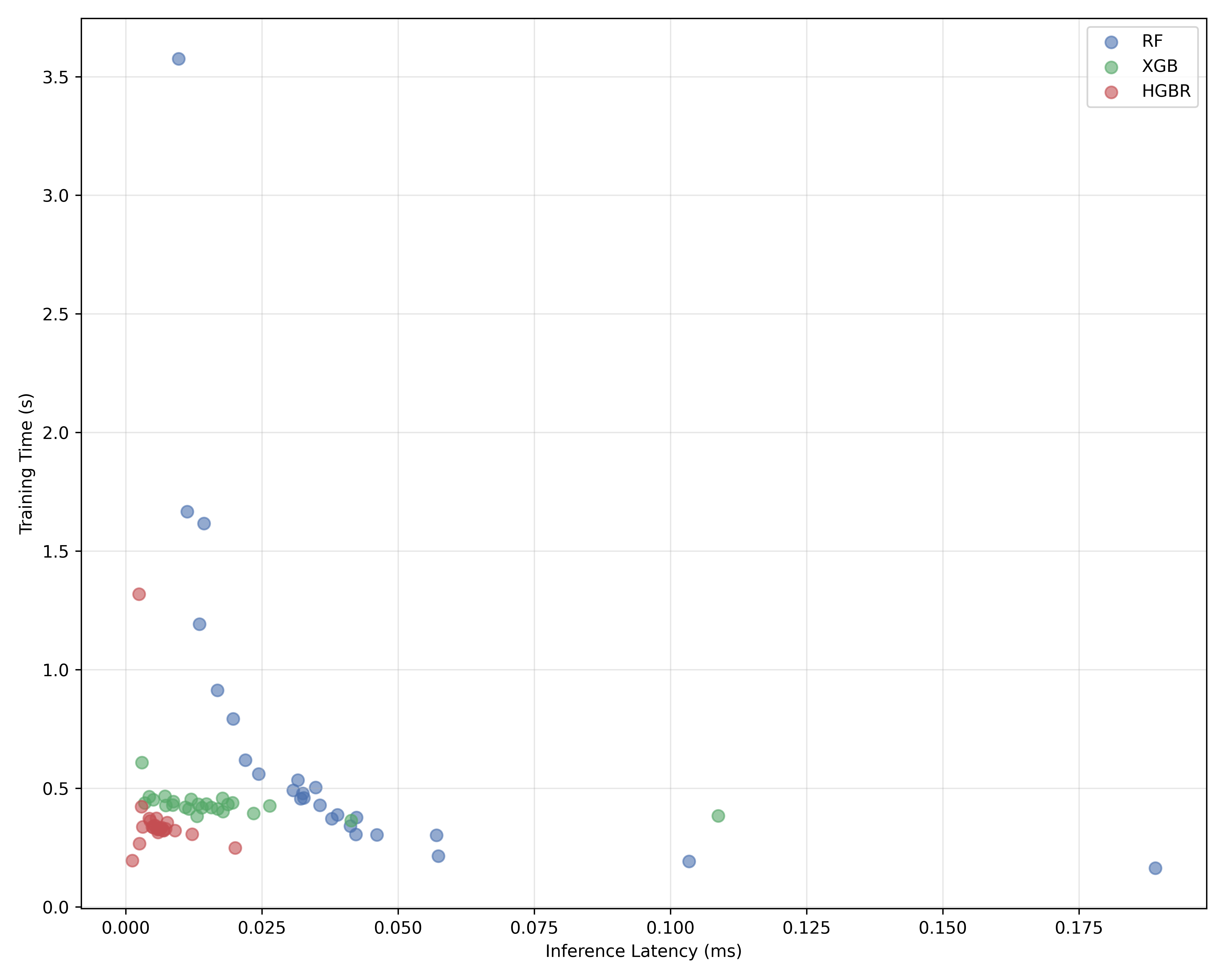}
\caption{Performance complexity analysis plotting inference latency versus training time across all model domain combinations.}
\label{fig:complexity}
\end{figure}

Figure~\ref{fig:complexity} plots inference latency versus training time for all model domain combinations. RF models form a scattered cluster in the high-latency, high-training-time region, while HGBR models tightly cluster in the bottom-left corner, confirming sub-millisecond inference and markedly reduced training times; the Canopy HGBR expert achieves the absolute minimum inference latency of 0.00122 ms. The universal XGB model offers a middle ground, with an inference latency of 0.00298 ms and a training time of 0.61 s. Section~\ref{sec:discussion} examines the theoretical basis for this latency hierarchy.

\subsection{TMAS Agentic Routing Proof of Concept}
To demonstrate the feasibility of the proposed multi-agent routing, we implemented a proof-of-concept simulation following the routing stage of Algorithm~\ref{alg:tmas}. Using the trained models, we constructed a registry of 23 Domain Micro-Agents covering all operator-mobility-traffic profile combinations; the remaining combinations default to the universal HGBR fallback. A rule-based Master Agent identifies the context of each incoming sample and routes it through the appropriate Operator Agent to the corresponding micro-agent, with each micro-agent using the architecture that achieved the highest $R^2$ for its domain during training.

\begin{figure*}[t]
\centering
\includegraphics[width=\textwidth]{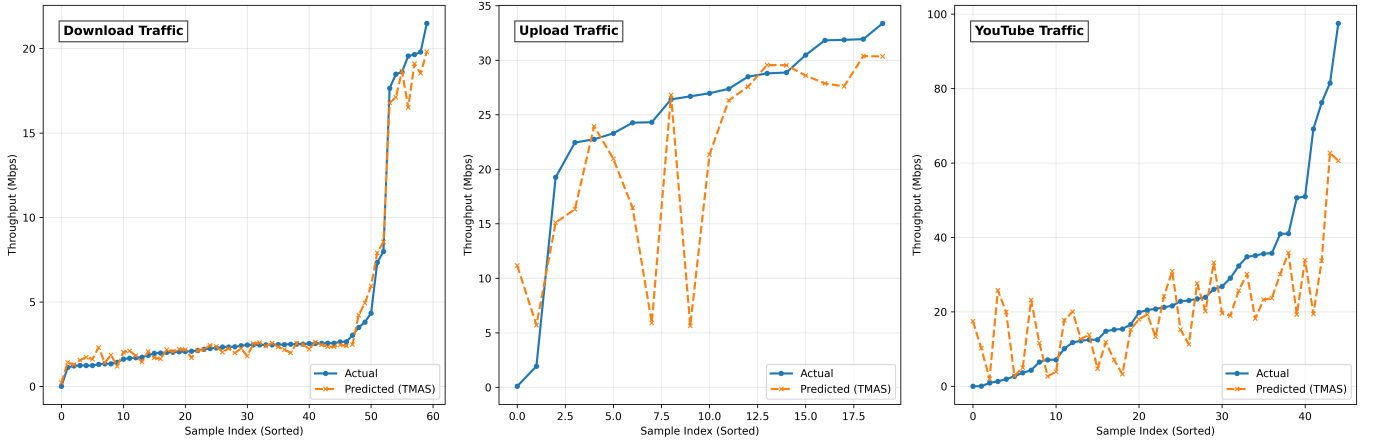}
\caption{Actual versus predicted throughput across traffic profiles using TMAS agentic routing.}
\label{fig:actual_pred}
\end{figure*}

Table~\ref{tab:routing_demo} reports 9 sample predictions with per-sample routing overhead, inference latency, and total end-to-end latency. Figure~\ref{fig:actual_pred} compares actual and predicted throughput: the Download trace shows a stable baseline with a sharp step-change spike at sample 50 that the TMAS micro-agent tracks closely, while the YouTube trace matches the low-throughput baseline but misses the pronounced spike after sample 40, consistent with the ABR stochasticity gap examined in Section~\ref{sec:discussion}.

\begin{figure*}[t]
\centering
\includegraphics[width=\textwidth]{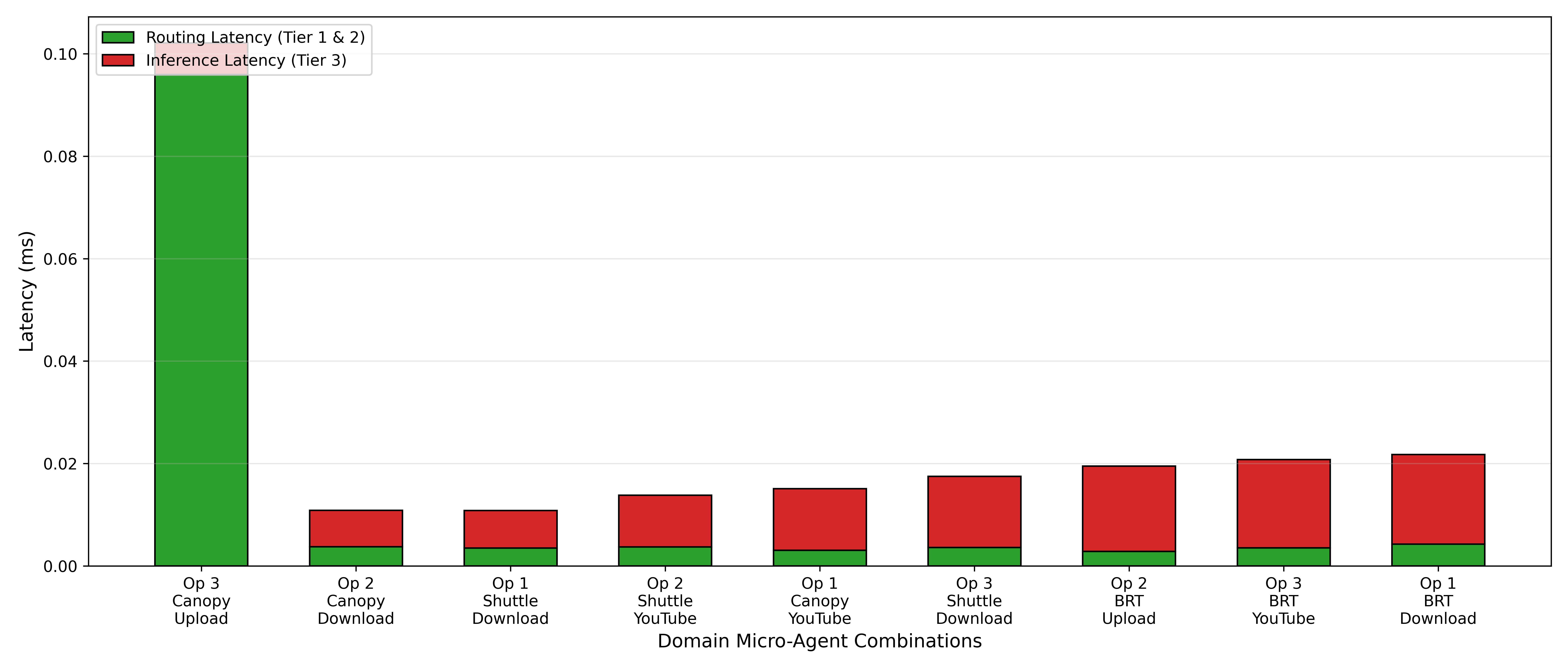}
\caption{Latency breakdown showing Tier 1 and Tier 2 routing overhead versus Tier 3 inference latency across sampled predictions.}
\label{fig:latency_breakdown}
\end{figure*}

Figure~\ref{fig:latency_breakdown} visualizes this end-to-end latency as a stacked bar chart. Most combinations complete with negligible latency bars, but the Operator 3 Canopy Upload sample, corresponding to the final row of Table~\ref{tab:routing_demo}, incurs a routing latency of 0.12630 ms. Here the Master Agent finds no dedicated micro-agent for this data-sparse context and routes to the Universal Fallback model instead, demonstrating that the routing architecture degrades gracefully rather than failing when training data are insufficient.

\begin{table}[ht]
\caption{Routing Demonstration: Sample Predictions with End-to-End Latency}
\label{tab:routing_demo}
\centering
\footnotesize
\setlength{\tabcolsep}{4pt}
\begin{tabular}{@{}cccccrrr@{}}
\toprule
\textbf{Actual} & \textbf{Pred.} & \textbf{Op.} & \textbf{Mob.} & \textbf{Use Case} & \textbf{Route} & \textbf{Inf.} & \textbf{Total} \\
(Mbps) & (Mbps) & & & & (ms) & (ms) & (ms) \\ \midrule
2.55   & 2.11   & Op.~1 & BRT      & Download & 0.00380 & 0.01749 & 0.02129 \\
17.06  & 17.20  & Op.~1 & Shuttle  & Download & 0.00410 & 0.00735 & 0.01145 \\
137.55 & 100.79 & Op.~1 & Canopy   & YouTube  & 0.00550 & 0.01202 & 0.01752 \\
27.67  & 26.11  & Op.~2 & BRT      & Upload   & 0.00460 & 0.01665 & 0.02125 \\
16.52  & 4.61   & Op.~2 & Shuttle  & YouTube  & 0.00510 & 0.01008 & 0.01518 \\
1.26   & 1.60   & Op.~2 & Canopy   & Download & 0.00520 & 0.00711 & 0.01231 \\
39.27  & 40.33  & Op.~3 & BRT      & YouTube  & 0.00470 & 0.01725 & 0.02195 \\
2.50   & 2.04   & Op.~3 & Shuttle  & Download & 0.00410 & 0.01388 & 0.01798 \\
24.31  & 5.90   & \multicolumn{3}{c}{Universal Fallback [HGBR]} & 0.12630 & 0.00617 & 0.13247 \\
\bottomrule
\end{tabular}
\end{table}

\section{Discussion}
\label{sec:discussion}
The results consistently reveal three distinct sources of variability in prediction performance, with significant implications for deploying machine learning models in urban 5G networks. First, the pronounced contrast between the predictability of download traffic ($R^2 \approx 0.931$) and YouTube streaming ($R^2 \approx 0.125$ to $0.339$) confirms that application layer logic fundamentally alters the relationship between radio conditions and achievable throughput, as evidenced by the horizontal dispersion in the YouTube scatter plot (Figure~\ref{fig:scatter}) and the missed spike in the line graph (Figure~\ref{fig:actual_pred}); the mechanism underlying this gap is examined in detail below.

Second, the per-operator per-mobility results establish a clear hierarchy of predictability that is consistent across operators. As shown in Table~\ref{tab:domain_models}, the Canopy Walk environment is the most consistently challenging, achieving at most $R^2 = 0.335$ (Operator 1, RF). For Operators 2 and 3, BRT outperforms Shuttle Bus, which in turn outperforms Canopy, reflecting increasing multipath complexity, handover frequency, and shadowing severity from elevated guideways to street level to pedestrian foliage. The anomalously high Shuttle Bus performance for Operator 1 ($R^2 = 0.924$) is driven by traffic composition rather than radio conditions, underscoring that model performance is jointly determined by the physical environment and the statistical distribution of traffic profiles within each domain, further motivating TMAS's silo isolation strategy since a model that performs well in one operator's domain cannot be assumed to generalize to another.

Third, the per-operator per-traffic results reveal a substantial variance in download predictability across operators, from $R^2 = 0.931$ (Operator 1) to $R^2 = 0.805$ (Operator 3). Reduced sample density for uplink traffic in a subset of operator-mobility combinations (Table~\ref{tab:data_stats}) likely contributes to the lower overall predictability of the upload models relative to the download profile; the routing proof of concept in Section~\ref{sec:results} demonstrates the practical consequence of this sparsity, where the Operator 3 Canopy Upload context lacks a dedicated micro-agent and is routed instead to the heavier Universal Fallback model. This variation likely stems from differences in tower density, backhaul capacity, and scheduling policies, so a model trained on one carrier's data may be unreliable for a roaming device or multi-operator virtual network. Federated learning approaches \cite{10838521, 10.1145/3709141} could enable collaborative model training without exposing proprietary operator data; however, the non-i.i.d.\ nature of the operator silos identified here suggests that standard federated averaging would still suffer from weight divergence without context-aware personalization.

\subsection{Physical Interpretation of Feature Importance}
This dominance of geographic coordinates suggests the ensemble models implicitly construct a granular radio environment map, memorizing localized degradation zones such as where a shuttle bus enters a concrete underpass or canopy foliage is densest. Prioritizing the rolling RSRP average over its instantaneous counterpart indicates the models filter out transient, small-scale Rayleigh fading in favor of smoothed, large-scale shadowing trends, implying that the physical constraints of the Sunway City environment, rather than network configuration, are the ultimate arbiters of application throughput.

\subsection{The YouTube Stochasticity Gap as a Cross-Layer Problem}
The poor and highly variable performance on YouTube ($R^2 \approx 0.125$ to $0.339$) should not be interpreted as a model weakness but as empirical evidence of a fundamental stochasticity gap arising from buffer-channel decoupling. In ABR streaming, the device halts data requests when the playout buffer is full, even under excellent 5G conditions, and requests aggressively during buffer underflow despite degraded radio conditions. Throughput is therefore not monotonic in signal strength alone, but also depends on buffer occupancy and the ABR algorithm's bitrate ladder decisions. Accurate throughput prediction for streaming services requires cross-layer signaling that exposes application layer buffer state to the physical layer prediction model, telemetry that future 6G AI-native networks should incorporate into the agentic prediction pipeline to prevent the baseline-collapsed predictions observed here.

\subsection{Theoretical Complexity and Real-Time Feasibility}
The empirical latency measurements are consistent with theoretical complexity expectations. Random Forest evaluates all $M$ trees to depth $d$ for every prediction, giving $O(M \cdot d)$ inference complexity with a large constant due to memory-scattered tree traversal. HGBR and XGBoost share the same $O(T \cdot B)$ asymptotic complexity per boosting round $T$ over $B$ histogram bins; however, HGBR's contiguous bin table layout and XGBoost's cache-aware block structure reduce the memory access constant significantly. This explains why HGBR achieves the lowest universal inference latency (0.00246 ms) and the absolute minimum at the micro-agent level (0.00122 ms for the Canopy expert), while XGBoost achieves the shortest training time (0.61 s universal) due to its approximate greedy splitting.

\subsection{TMAS as a 6G Core Orchestration Engine}
While TMAS validation relies on device-side telemetry collected via Nemo Handy, this reflects a necessary perception layer rather than a data constraint: network-side monitoring has limited visibility into application layer states, explaining why the YouTube evaluations collapse to $R^2 \approx 0.125$ under radio conditions that yield $R^2 = 0.931$ for download, since the network cannot allocate resources for a streaming session without knowing the application's buffer state. TMAS is therefore designed for deployment within multi-access edge computing servers or as an xApp within the Near Real-Time RAN Intelligent Controller (Near-RT RIC) of an O-RAN architecture, where the Master Agent ingests lightweight, anonymized state vectors comprising mobility mode and traffic profile via the minimization of drive tests reporting framework in 3GPP TS~37.320 \cite{3gpp_ts_37_320_rel18} and the network data analytics function in 3GPP TS~23.288 \cite{3gpp_ts_23_288_rel18}, carried over the O-RAN E2 interface connecting the E2 nodes (gNB-CU/DU) to the Near-RT RIC. Because this telemetry is inherently operator-specific, the TMAS Operator Agent silo structure maps naturally onto existing O-RAN multi-operator boundaries, circumventing weight divergence without inter-operator data sharing.

With inference latency as low as 0.00122 ms, TMAS terminal nodes execute predictions well within the 10 ms Near-RT RIC control loop budget, leaving ample margin for routing overhead (0.004 ms to 0.006 ms per sample for dedicated micro-agent routes, rising to 0.126 ms for the Universal Fallback case, Table~\ref{tab:routing_demo}) and actuator signaling, enabling the RIC to proactively trigger network slicing, predictive handover, or edge caching before buffer underflow occurs. Short micro-agent training times further enable frequent model refresh cycles without the retraining overhead of monolithic deep learning architectures. Closing the ABR stochasticity gap would additionally require incorporating buffer state telemetry into the micro-agent feature set via the same minimization of drive tests extensions \cite{3gpp_ts_37_320_rel18}, allowing the RIC to distinguish genuine radio degradation from ABR buffer saturation, transitioning throughput prediction from a reactive, localized measurement to a proactive, network-wide control variable.

\section{Conclusion}
This study has presented a comprehensive multi-domain framework for 5G NR throughput prediction within the high-entropy environment of Sunway City, quantifying the severe generalization limits of monolithic predictive modeling and proposing TMAS to address them.

The empirical findings validate this architectural shift across three dimensions. First, TMAS Domain Micro-Agents consistently outperform monolithic models, achieving $R^2 = 0.931$ for download traffic (Operator 1, RF) by isolating context-specific channel behaviors that a universal model cannot capture. Second, the cross-domain evaluation reveals a severe predictability ceiling for ABR-based YouTube streaming ($R^{2}\approx 0.125$ to $0.339$), establishing a fundamental cross-layer stochasticity gap that physical layer agents alone cannot bridge. Third, TMAS terminal nodes achieve inference latency as low as 0.00122 ms (Canopy, HGBR) and rapid training times, with total end-to-end routing latency remaining well below 0.5 ms per sample, comfortably satisfying the 10 ms Near-RT RIC control budget.

These results confirm the three contributions outlined in Section I. Overcoming the urban predictability bottleneck requires abandoning universal modeling in favor of dynamic, agentic routing; network designers should consider context-aware, operator-specific micro-agents as a building block for future intelligent network controllers, and should incorporate application-layer buffer states into the prediction pipeline, moving toward more AI-native 6G network orchestration.

Future work will extend the measurement campaign to multiple cities and standalone 5G and emerging 6G testbeds, targeting sparsely populated regions of the operator-mobility-traffic matrix; incorporate application-layer telemetry such as video buffer occupancy to close the ABR stochasticity gap; and explore lightweight federated personalization (e.g., clustered federated learning or partial parameter sharing) alongside quantized or pruned HGBR variants for ultra-low-power edge deployment.

\bibliographystyle{IEEEtran}
\bibliography{ref}

@Article{s19173651,
AUTHOR = {Adel Aly, Ahmed and M. ELAttar, Hussein and ElBadawy, Hesham and Abbas, Wael},
TITLE = {Aggregated Throughput Prediction for Collated Massive Machine-Type Communications in 5G Wireless Networks},
JOURNAL = {Sensors},
VOLUME = {19},
YEAR = {2019},
NUMBER = {17},
ARTICLE-NUMBER = {3651},
URL = {https://www.mdpi.com/1424-8220/19/17/3651},
PubMedID = {31443468},
ISSN = {1424-8220},
DOI = {10.3390/s19173651}}

@ARTICLE{11278580,
  author={Dabarera, Sahan and Pochaba, Sabrina and Maier, Christian and Herlich, Matthias and Dorfinger, Peter},
  journal={IEEE Access}, 
  title={Explaining and Predicting Mobile Network Performance From Multi-Operator Data Using Machine Learning}, 
  year={2025},
  volume={13},
  number={},
  pages={208788-208800},
  doi={10.1109/ACCESS.2025.3640944}}

@ARTICLE{9495144,
  author={Minovski, Dimitar and Ögren, Niclas and Mitra, Karan and Åhlund, Christer},
  journal={IEEE Transactions on Mobile Computing}, 
  title={Throughput Prediction Using Machine Learning in LTE and 5G Networks}, 
  year={2023},
  volume={22},
  number={3},
  pages={1825-1840},
  doi={10.1109/TMC.2021.3099397}}

@ARTICLE{11170415,
  author={Zhao, Jintang and Cheng, Jianfeng and Li, Shuo and Liu, Guanghua and Yang, Liuchang and Jiang, Tao},
  journal={IEEE Transactions on Network Science and Engineering}, 
  title={Mobile Network Optimization in High-Speed Railway Tunnels: An Adaptive Handover Based on Train Position and Signal Quality Prediction}, 
  year={2026},
  volume={13},
  number={},
  pages={2251-2269},
  doi={10.1109/TNSE.2025.3611587}}

@inproceedings{10.1145/3636534.3694725,
author = {Feng, Zihao and Chen, Xingyu and Cao, Xuyang and Zhang, Xinyu},
title = {Hybrid Data-Driven and Simulation-Driven Prediction of mmWave Network Performance},
year = {2024},
isbn = {9798400704895},
publisher = {Association for Computing Machinery},
address = {New York, NY, USA},
url = {https://doi.org/10.1145/3636534.3694725},
doi = {10.1145/3636534.3694725},
booktitle = {Proceedings of the 30th Annual International Conference on Mobile Computing and Networking},
pages = {1998–2003},
numpages = {6},
location = {Washington D.C., DC, USA},
series = {ACM MobiCom '24}
}

@ARTICLE{10147378,
  author={Jung, Jewon and Lee, Sugi and Shin, Jaemin and Kim, Yusung},
  journal={IEEE Internet of Things Journal}, 
  title={Self-Attention-Based Uplink Radio Resource Prediction in 5G Dual Connectivity}, 
  year={2023},
  volume={10},
  number={22},
  pages={19925-19936},
  doi={10.1109/JIOT.2023.3283490}}

@article{HELMY2025104324,
title = {Autoformer-based mobility and handoff-aware prediction for QoE enhancement in adaptive video streaming in 4G/5G networks},
journal = {Journal of Network and Computer Applications},
volume = {243},
pages = {104324},
year = {2025},
issn = {1084-8045},
doi = {https://doi.org/10.1016/j.jnca.2025.104324},
url = {https://www.sciencedirect.com/science/article/pii/S1084804525002218},
author = {Maram Helmy and Mohamed S. Hassan and Mahmoud H. Ismail and Usman Tariq}
}

@ARTICLE{10770241,
  author={An, Congkai and Zhang, Huanhuan and Kang, Jingyang and Liu, Zhuo and Zhou, Anfu and Liu, Liang and Ma, Huadong},
  journal={IEEE Transactions on Circuits and Systems for Video Technology}, 
  title={Enhancing QoE of Adaptive Video Streaming by Generating Fine-Grained Throughput}, 
  year={2025},
  volume={35},
  number={4},
  pages={3853-3866},
  doi={10.1109/TCSVT.2024.3508262}}

@Article{electronics11081227,
AUTHOR = {Eyceyurt, Engin and Egi, Yunus and Zec, Josko},
TITLE = {Machine-Learning-Based Uplink Throughput Prediction from Physical Layer Measurements},
JOURNAL = {Electronics},
VOLUME = {11},
YEAR = {2022},
NUMBER = {8},
ARTICLE-NUMBER = {1227},
URL = {https://www.mdpi.com/2079-9292/11/8/1227},
ISSN = {2079-9292},
DOI = {10.3390/electronics11081227}
}

@article{10.1145/3724400,
author = {Raca, Darijo and Provan, Gregory and Zahran, Ahmed},
title = {M2ATURE: Mobile Multistage Throughput Prediction for Adaptive Video Streaming in Cellular Networks},
year = {2026},
issue_date = {March 2026},
publisher = {Association for Computing Machinery},
address = {New York, NY, USA},
volume = {22},
number = {3},
issn = {1551-6857},
url = {https://doi.org/10.1145/3724400},
doi = {10.1145/3724400},
journal = {ACM Trans. Multimedia Comput. Commun. Appl.},
month = feb,
articleno = {64},
numpages = {17}
}

@inproceedings{10.1145/3651863.3651878,
author = {Nolan, Killian and Raca, Darijo and Provan, Gregory and Zahran, Ahmed},
title = {MATURE: Multistage Throughput Prediction for Adaptive Video Streaming in Cellular Networks},
year = {2024},
isbn = {9798400706134},
publisher = {Association for Computing Machinery},
address = {New York, NY, USA},
url = {https://doi.org/10.1145/3651863.3651878},
doi = {10.1145/3651863.3651878},
booktitle = {Proceedings of the 34th Workshop on Network and Operating System Support for Digital Audio and Video},
pages = {15–21},
numpages = {7},
location = {Bari, Italy},
series = {NOSSDAV '24}
}

@inproceedings{10.1145/3651890.3672250,
author = {Ye, Wei and Hu, Xinyue and Sleder, Steven and Zhang, Anlan and Dayalan, Udhaya Kumar and Hassan, Ahmad and Fezeu, Rostand A. K. and Jajoo, Akshay and Lee, Myungjin and Ramadan, Eman and Qian, Feng and Zhang, Zhi-Li},
title = {Dissecting Carrier Aggregation in 5G Networks: Measurement, QoE Implications and Prediction},
year = {2024},
isbn = {9798400706141},
publisher = {Association for Computing Machinery},
address = {New York, NY, USA},
url = {https://doi.org/10.1145/3651890.3672250},
doi = {10.1145/3651890.3672250},
booktitle = {Proceedings of the ACM SIGCOMM 2024 Conference},
pages = {340–357},
numpages = {18},
location = {Sydney, NSW, Australia},
series = {ACM SIGCOMM '24}
}

@ARTICLE{10950386,
  author={Lin, Fan-Hao and Huang, Tzu-Hao and Wen, Chao-Kai and Duong, Trung Q.},
  journal={IEEE Wireless Communications Letters}, 
  title={Geo2ComMap: Deep Learning-Based MIMO Throughput Prediction Using Geographic Data}, 
  year={2025},
  volume={14},
  number={6},
  pages={1831-1835},
  doi={10.1109/LWC.2025.3558198}}

@ARTICLE{10342799,
  author={Biswas, Mayukh and Chakraborty, Ayan and Palit, Basabdatta},
  journal={IEEE Transactions on Vehicular Technology}, 
  title={A Kalman Filter Based Low Complexity Throughput Prediction Algorithm for 5G Cellular Networks}, 
  year={2024},
  volume={73},
  number={5},
  pages={7089-7101},
  doi={10.1109/TVT.2023.3339649}}

@ARTICLE{11059833,
  author={Sayeed, Rayyan and Miller, Raymond and Sayeed, Zulfiquar},
  journal={Journal of Cyber Security and Mobility}, 
  title={Throughput Prediction Across Heterogeneous Boundaries in Wireless Communications}, 
  year={2015},
  volume={4},
  number={4},
  pages={233-258},
  doi={10.13052/jcsm2245-1439.441}}

@article{Al-Thaedan2023,
  author    = {Abbas Al-Thaedan and Zaenab Shakir and Ahmed Yaseen Mjhool and Ruaa Alsabah and Ali Al-Sabbagh and Monera Salah and Josko Zec},
  title     = {Downlink throughput prediction using machine learning models on 4G-LTE networks},
  journal   = {International Journal of Information Technology},
  year      = {2023},
  volume    = {15},
  number    = {6},
  pages     = {2987--2993},
  doi       = {10.1007/s41870-023-01358-9},
  url       = {https://doi.org/10.1007/s41870-023-01358-9},
  issn      = {2511-2112}
}

@INPROCEEDINGS{10914824,
  author={Dhanalakshmi, K. S. and Deepak, Kambati Lakshmi and Chandu, Chinnam Sai and Reddy, Palam Venkata Krishna and Reddy, Nagireddy Vishnu Vardhan and Manikanta, Kesanasetty Naga},
  booktitle={2025 3rd International Conference on Intelligent Data Communication Technologies and Internet of Things (IDCIoT)}, 
  title={Real-Time Traffic Load Prediction in Cellular Networks Using Cutting-Edge Machine Learning Approaches}, 
  year={2025},
  volume={},
  number={},
  pages={572-577},
  doi={10.1109/IDCIOT64235.2025.10914824}}

@ARTICLE{9878077,
  author={Li, Lanlan and Ye, Tao},
  journal={Intelligent and Converged Networks}, 
  title={Research on throughput prediction of 5G network based on LSTM}, 
  year={2022},
  volume={3},
  number={2},
  pages={217-227},
  doi={10.23919/ICN.2022.0006}}

@article{LI2025111669,
title = {A hybrid spatiotemporal LSTM and transformer network for cellular traffic prediction},
journal = {Computer Networks},
volume = {272},
pages = {111669},
year = {2025},
issn = {1389-1286},
doi = {https://doi.org/10.1016/j.comnet.2025.111669},
url = {https://www.sciencedirect.com/science/article/pii/S138912862500636X},
author = {A-Min Li and Jingqi Li}
}

@article{10.1145/3703629,
author = {Lv, Jiamei and Lin, Yuxiang and Hou, Mingxin and Li, Yeming and Gao, Yi and Dong, Wei},
title = {Accurate Bandwidth and Delay Prediction for 5G Cellular Networks},
year = {2025},
issue_date = {May 2025},
publisher = {Association for Computing Machinery},
address = {New York, NY, USA},
volume = {25},
number = {2},
issn = {1533-5399},
url = {https://doi.org/10.1145/3703629},
doi = {10.1145/3703629},
journal = {ACM Trans. Internet Technol.},
month = apr,
articleno = {8},
numpages = {25}
}

@INPROCEEDINGS{11167476,
  author={Waheed, Mohamed Tharwat and Khattab, Ahmed and Fahmy, Yasmine},
  booktitle={2025 Intelligent Methods, Systems, and Applications(IMSA)}, 
  title={Downlink Throughput Prediction In Cellular Networks Using Multimodal Deep Learning Model}, 
  year={2025},
  volume={},
  number={},
  pages={334-339},
  doi={10.1109/IMSA65733.2025.11167476}}

@INPROCEEDINGS{11274662,
  author={Maurya, Poonam and Ramírez-Arroyo, Alejandro and Sørensen, Troels Bundgaard},
  booktitle={2025 IEEE 36th International Symposium on Personal, Indoor and Mobile Radio Communications (PIMRC)}, 
  title={Network KPI Prediction in Mobile Networks Using CNN-LSTM Time Series Model}, 
  year={2025},
  volume={},
  number={},
  pages={1-6},
  doi={10.1109/PIMRC62392.2025.11274662}}

@INPROCEEDINGS{11317760,
  author={V M, Sapna and S S, Achyuth and H.B, Prasad},
  booktitle={2025 IEEE Future Networks World Forum (FNWF)}, 
  title={Throughput Prediction in Dense WLANs Using TabTransformer Models Trained on Simulated Scenarios}, 
  year={2025},
  volume={},
  number={},
  pages={1-4},
  doi={10.1109/FNWF66845.2025.11317760}}

@INPROCEEDINGS{10992709,
  author={Ni, Nan and Fujii, Takeo},
  booktitle={2025 International Conference on Information Networking (ICOIN)}, 
  title={Random Forest Prediction of WLAN Throughput using Communication Logs and Channel Occupancy Rate}, 
  year={2025},
  volume={},
  number={},
  pages={191-196},
  doi={10.1109/ICOIN63865.2025.10992709}}

@ARTICLE{10971899,
  author={Zhang, Peiyun and Fan, Jiajun and Chen, Yutong and Huang, Wenjun and Zhu, Haibin and Zhao, Qinglin},
  journal={IEEE Transactions on Services Computing}, 
  title={An End-to-End Deep Learning QoS Prediction Model Based on Temporal Context and Feature Fusion}, 
  year={2025},
  volume={18},
  number={3},
  pages={1232-1244},
  doi={10.1109/TSC.2025.3562324}}

@ARTICLE{11141395,
  author={Yau, Shuaibu and Awiphan, Suphakit and Bootkrajang, Jakramate and Katto, Jiro},
  journal={IEEE Access}, 
  title={A Robust Throughput Estimation in Edge-Assisted Adaptive Bitrate Streaming Networks}, 
  year={2025},
  volume={13},
  number={},
  pages={152598-152607},
  doi={10.1109/ACCESS.2025.3602651}}

@ARTICLE{10056411,
  author={Palit, Basabdatta and Sen, Argha and Mondal, Abhijit and Zunaid, Ayan and Jayatheerthan, Jay and Chakraborty, Sandip},
  journal={IEEE Transactions on Network and Service Management}, 
  title={Improving UE Energy Efficiency Through Network-Aware Video Streaming Over 5G}, 
  year={2023},
  volume={20},
  number={3},
  pages={3487-3500},
  doi={10.1109/TNSM.2023.3250520}}

@ARTICLE{10192432,
  author={Sultan, Mohamad T. and El Sayed, Hesham},
  journal={IEEE Access}, 
  title={QoE-Aware Analysis and Management of Multimedia Services in 5G and Beyond Heterogeneous Networks}, 
  year={2023},
  volume={11},
  number={},
  pages={77679-77688},
  doi={10.1109/ACCESS.2023.3298556}}

@ARTICLE{10838521,
  author={Nugroho, Kukuh and Hendrawan and Iskandar},
  journal={IEEE Access}, 
  title={Comparative Analysis of Federated and Centralized Learning Systems in Predicting Cellular Downlink Throughput Using CNN}, 
  year={2025},
  volume={13},
  number={},
  pages={22745-22763},
  doi={10.1109/ACCESS.2025.3528527}}

@article{10.1145/3709141,
author = {Zou, Guobing and Lin, Shiyi and Wu, Shaogang and Hu, Shengxiang and Yang, Song and Gan, Yanglan and Zhang, Bofeng and Chen, Yixin},
title = {Combining Personalized Federated Hypernetworks and Shared Residual Learning for Distributed QoS Prediction},
year = {2025},
issue_date = {September 2025},
publisher = {Association for Computing Machinery},
address = {New York, NY, USA},
volume = {20},
number = {3},
issn = {1556-4665},
url = {https://doi.org/10.1145/3709141},
doi = {10.1145/3709141},
journal = {ACM Trans. Auton. Adapt. Syst.},
month = sep,
articleno = {23},
numpages = {25}
}

@article{RATHORE2026111824,
title = {Throughput optimization of cognitive radio networks in high-traffic driven ecosystems with threshold selection technique},
journal = {Computer Networks},
volume = {274},
pages = {111824},
year = {2026},
issn = {1389-1286},
doi = {https://doi.org/10.1016/j.comnet.2025.111824},
url = {https://www.sciencedirect.com/science/article/pii/S138912862500790X},
author = {Anmol Shalom Rathore and Alok Kumar}
}

@ARTICLE{10929655,
  author={Afshar, Sepideh and Razavi, Reza and Moshirpour, Mohammad},
  journal={IEEE Transactions on Machine Learning in Communications and Networking}, 
  title={Closed-Loop Clustering-Based Global Bandwidth Prediction in Real-Time Video Streaming}, 
  year={2025},
  volume={3},
  number={},
  pages={448-462},
  doi={10.1109/TMLCN.2025.3551689}}

@ARTICLE{10007850,
  author={Fauzi, Mohd Fazuwan Ahmad and Nordin, Rosdiadee and Abdullah, Nor Fadzilah and Alobaidy, Haider A. H. and Behjati, Mehran},
  journal={IEEE Access}, 
  title={Machine Learning-Based Online Coverage Estimator (MLOE): Advancing Mobile Network Planning and Optimization}, 
  year={2023},
  volume={11},
  number={},
  pages={3096-3109},
  doi={10.1109/ACCESS.2023.3234566}}

@ARTICLE{10418223,
  author={Yuliana, Hajiar and Iskandar and Hendrawan},
  journal={IEEE Access}, 
  title={Comparative Analysis of Machine Learning Algorithms for 5G Coverage Prediction: Identification of Dominant Feature Parameters and Prediction Accuracy}, 
  year={2024},
  volume={12},
  number={},
  pages={18939-18956},
  doi={10.1109/ACCESS.2024.3361403}}

@article{HASAN2026100721,
title = {Integrating agentic AI and digital twins for intelligent decision-making systems},
journal = {Array},
volume = {29},
pages = {100721},
year = {2026},
issn = {2590-0056},
doi = {https://doi.org/10.1016/j.array.2026.100721},
url = {https://www.sciencedirect.com/science/article/pii/S2590005626000445},
author = {Agus Hasan and Dong Trong Nguyen}
}

@techreport{3gpp_ts_37_320_rel18,
  author      = {{3GPP}},
  title       = {{Radio Measurement Collection for Minimization of Drive Tests (MDT); Overall Description; Stage 2}},
  institution = {3rd Generation Partnership Project (3GPP)},
  type        = {Technical Specification (TS)},
  number      = {37.320},
  month       = feb,
  year        = {2026},
  url         = {https://portal.3gpp.org/desktopmodules/Specifications/SpecificationDetails.aspx?specificationId=2602}
}

@techreport{3gpp_ts_23_288_rel18,
  author      = {{3GPP}},
  title       = {{Architecture Enhancements for 5G System (5GS) to Support Network Data Analytics Services}},
  institution = {3rd Generation Partnership Project (3GPP)},
  type        = {Technical Specification (TS)},
  number      = {23.288},
  month       = jan,
  year        = {2026},
  url         = {https://portal.3gpp.org/desktopmodules/Specifications/SpecificationDetails.aspx?specificationId=3579}
}

@dataset{kabeer2026multioperator,
  doi = {10.17632/8VVN3K3HMZ},
  url = {https://data.mendeley.com/datasets/8vvn3k3hmz},
  author = {Kabeer, Muhammad and Nuriftitah, Nadiva and Nordin, Rosdiadee},
  title = {Multi-Operator Dataset for Throughput Prediction Across Diverse Mobility Modes},
  publisher = {Mendeley Data},
  year = {2026}
}

\end{document}